\documentclass{article}
\usepackage{spconf,amsmath,graphicx}

\title{Color Deconvolution applied to Domain Adaptation in HER2 histopathological images}
\name{David Anglada-Rotger, Ferran Marques, Montse Pardas}
\address{Image Processing Group, Universitat Politècnica de Catalunya, Barcelona, Spain}
\begin{document}
%
\maketitle
\begin{abstract}
 Breast cancer early detection is crucial for improving patient outcomes. The Institut Catal\`a de la Salut (ICS) has launched the DigiPatICS project to develop and implement artificial intelligence algorithms to assist with the diagnosis of cancer. In this paper, we propose a new approach for facing the color normalization problem in HER2-stained histopathological images of breast cancer tissue, posed as an style transfer problem. We combine the Color Deconvolution technique with the Pix2Pix GAN network to present a novel approach to correct the color variations between different HER2 stain brands. Our approach focuses on maintaining the HER2 score of the cells in the transformed images, which is crucial for the HER2 analysis. Results demonstrate that our final model outperforms the state-of-the-art image style transfer methods in maintaining the cell classes in the transformed images and is as effective as them in generating realistic images.
\end{abstract}

\section{Introduction}\label{sec:intro}

Cancer is a major public health concern worldwide, with more than 10 million deaths each year. Early detection of cancer is crucial for improving patient outcomes, and biopsy is the primary method for diagnosis. However, the process of diagnosing cancer can be time-consuming and subject to variability among hospitals and pathologists.

\begin{figure}[tb]
    \centering
    \includegraphics[scale=0.2]{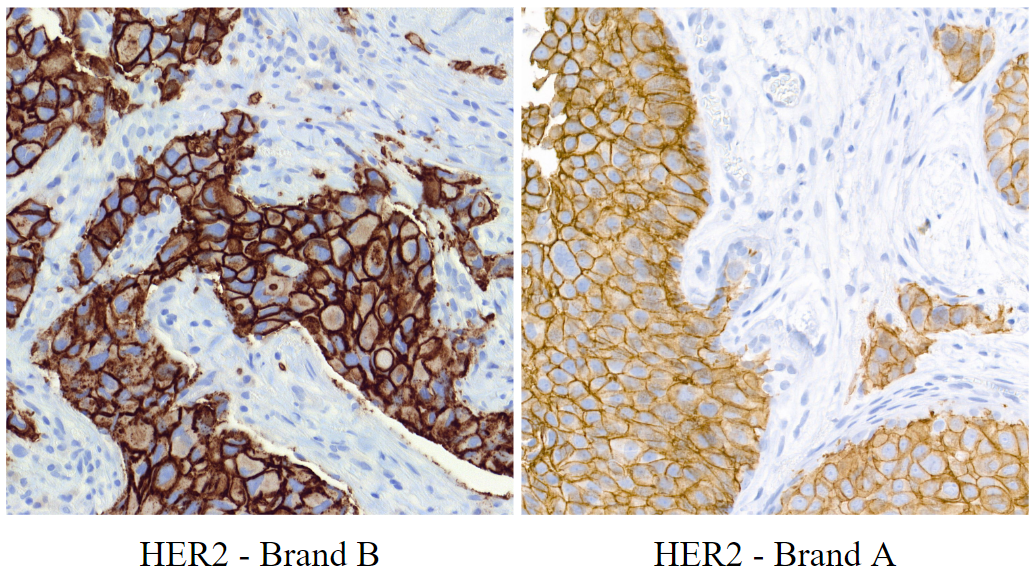}
    \caption{Example of the color variability between HER2 stain brands. HER2 tile from brand A (right) and brand B (left).}
    \label{fig:ex-var}
\end{figure}

The HER2 stain identifies the overproduction of the HER2 protein, a receptor in breast cells that controls cell growth and division. The HER2 stain classifies cells into four different classes based on the intensity and completeness of the stained membrane, and estimates the percentage of cells in each class to assign a score to the patient. However, HER2 stained images can vary significantly among different hospitals because of the use of different stain brands (Figure \ref{fig:ex-var}), leading to inconsistencies in diagnosis and treatment decisions. To address this challenge, we propose a novel approach to normalize color variations between HER2-stained histopathological images from different domains. Our approach combines the Color Deconvolution technique and the Pix2Pix GAN network to perform a style transfer that maintains the membrane intensity of each cell, a key feature used in cell classification for HER2 stained images. 

The ability to compute the translation between different HER2 stain brands is a crucial step in multicenter projects, as it enables the use of a single, common model for all hospitals regardless the use of different stain brands. This eliminates the need for acquiring separate datasets to train individual models for each strain brand, which can be a time-consuming and resource-intensive process. Moreover, this translation helps to ensure consistency in diagnosis and treatment decisions, ultimately improving patient outcomes. Our method outperforms in maintaining the cell classes, which is crucial in the context of HER2 stained images. Furthermore, our study demonstrates that the proposed approach is effective in generating images that are as realistic as those generated by state-of-the-art models.

\section{Related work}

\subsection{Color Deconvolution}\label{sec:color-dec}

The final result of the staining and scanning process are digital images called Whole Slide Images (WSIs). Colour deconvolution is an image processing technique that has been widely used in the analysis of WSIs. This technique, first defined by Ruifrok A. \textit{et al.} \cite{color-dec2}, enables the decomposition of an RGB image into channels representing the optical absorbance and transmittance of dyes. This is particularly useful in the case of tissue staining, where different dyes are used to highlight specific structures or features within the tissue. In the case of HER2 staining, Hematoxylin is used to stain nuclei blue, and the component called DAB stains the membrane brown.

Colour deconvolution \cite{color-dec} separates the different components of each stain, creating an image for each individual component. The original RGB image is orthonormal transformed, which enables the extraction of independent information about each stain contribution and the correct balancing of the absorption factor for each separate stain.

The algorithm converts the RGB image into the optical-density (OD) space. Each pure stain is characterized by a specific optical density for the light in each of the three RGB channels, which is represented by a $3 \times 1$ OD vector, describing the stain in the OD-converted RGB color space. Being $Y$ the OD space image, $n_s$ the number of stains and $n \times m$ the dimensions of the image, by defining the colour vector matrix $M \in \mathbf{R}^{3\times n_s}$ (composed by the normalized OD vectors) and the stain concentrations matrix $C \in \mathbf{R}^{n\times m\times n_s}$, we can decompose $Y$ in the following way: $Y^T = MC^T + N^T$. Here, $N$ just represents random noise. Having a reference matrix $M_{ref}$ for a specific stain brand, we can compute the Moore-Penrose pseudo-inverse of $M_{ref}$, $\hat{M}^{-1}$ \cite{color_dec_use}. Then, matrix $C$ can be estimated from $C = \hat{M}^{-1}Y$ and the color deconvolution of image $Y$ can be performed, decomposing it in terms of $C$ and $M$.


As explained in P\'erez-Bueno F. \textit{et al.} \cite{color_dec_use}, this image processing technique can directly be used to perform a domain adaptation between two different images. Being \(M_{ref}\) and \(C_{ref}\) the matrices from the reference domain image and \(M\) and \(C\) the matrices from the input image, the domain adaptation can be performed multiplying the reference color vector matrix \(M_{ref}\) times the input image stain concentration matrix \(C\). Using directly the Color Deconvolution technique to perform the domain adaptation is a linear operation. However, in the case of HER2 stain variability, the transformation from the stain brand A domain to the stain brand B domain is not a linear operation. Hence, in this work, we present a non-linear approach based on the color deconvolution operation that will solve the style transfer problem between the stain brands A and B in HER2 histopathological images.

\subsection{Generative Adversarial Networks (GANs)}

Generative Adversarial Networks (GANs), first presented by  Goodfellow I. \textit{et al.} \cite{gan}, have been widely used for various tasks in computer vision, including image generation, super-resolution, and style transfer. Pix2Pix GAN, first presented by Isola P. \textit{et al.} \cite{pix2pix}, is a type of GAN architecture that is used to generate images from a source domain to a target domain. Pix2Pix GANs are trained on paired images, where the source image is mapped to the target image, allowing the GAN to learn the mapping between the two domains.

As stated by Jose L. \textit{et al.} \cite{histo_gan}, GANs have achieved state-of-the-art results in many generation tasks, but also in various medical imaging applications such as medical image style tranfer, which is the aim of this work. In Shaban M. \textit{et al.} \cite{staingan}, they present StainGAN, a network inspired by CycleGAN that corrects the color variations in the stain process of Hematoxylin and Eosin (HE). Same problem has been adressed with GAN inspired achitectures by Zhou N. \textit{et al.} \cite{sota_1} and Cai S. \textit{et. al.} \cite{sota_2}. In Wollmann T. \textit{et al.} \cite{da_1}, the authors used a CycleGAN based architecture to transform the stained images in HE from different hospitals into data from one center to reduce variability and improve the classification step. In Ren J. \textit{et al.} \cite{da_2}, they face the same problem but with a different strategy: with two different datasets (source and target domains), they train the target network to extract the domain invariant features from the source input samples.

Overall, these works demonstrate the effectiveness of Pix2Pix GANs in correcting color variability among different WSI, by learning a mapping between the source and target domains. However, there is still a need for improving the performance and robustness of GAN-based color correction methods for histopathological images. In particular, as explained in Section \ref{sec:intro}, in the case of HER2 stain, the analysis of the stained image depends on the intensity of the stained membranes of the cells. Thus, it is crucial that the style transfer method does not alter this characteristic of each cell, in order to maintain their classes. As shown in Section \ref{sec:results}, GAN-based methods generate realistic images in the desired style but do not maintain the intensity of the membranes, altering the HER2 classes of the cells.

Hence, our contributions in this work are the following: (i) We present an approach that brings non-linearity to the Color Deconvolution technique, to deal with more complex color normalization problems, as the HER2 one. (ii) We present a GAN-based style transfer approach that, combined with the Color Deconvolution technique, ensures the maintenance of the membrane intensity of each cell, in order not to alter the HER2 score of the translated image.

\section{Dataset}

Training deep learning models on WSIs presents a significant challenge due to their large size. Algorithms developed for WSIs typically use tiles, which are cropped images of relevant zones with a size of $1024 \times 1024$ pixels.

It is also crucial to consider the various classes of cells present in HER2-stained WSIs. In order to ensure a balanced representation of these classes, 3-4 WSIs from patients in each domain were selected and classified according to defined HER2 scores, and tiles were extracted using the QuPath \cite{qpath} tool, a software platform for bioimage analysis.

\begin{figure*}[htb]
    \centering
    \includegraphics[scale=0.20]{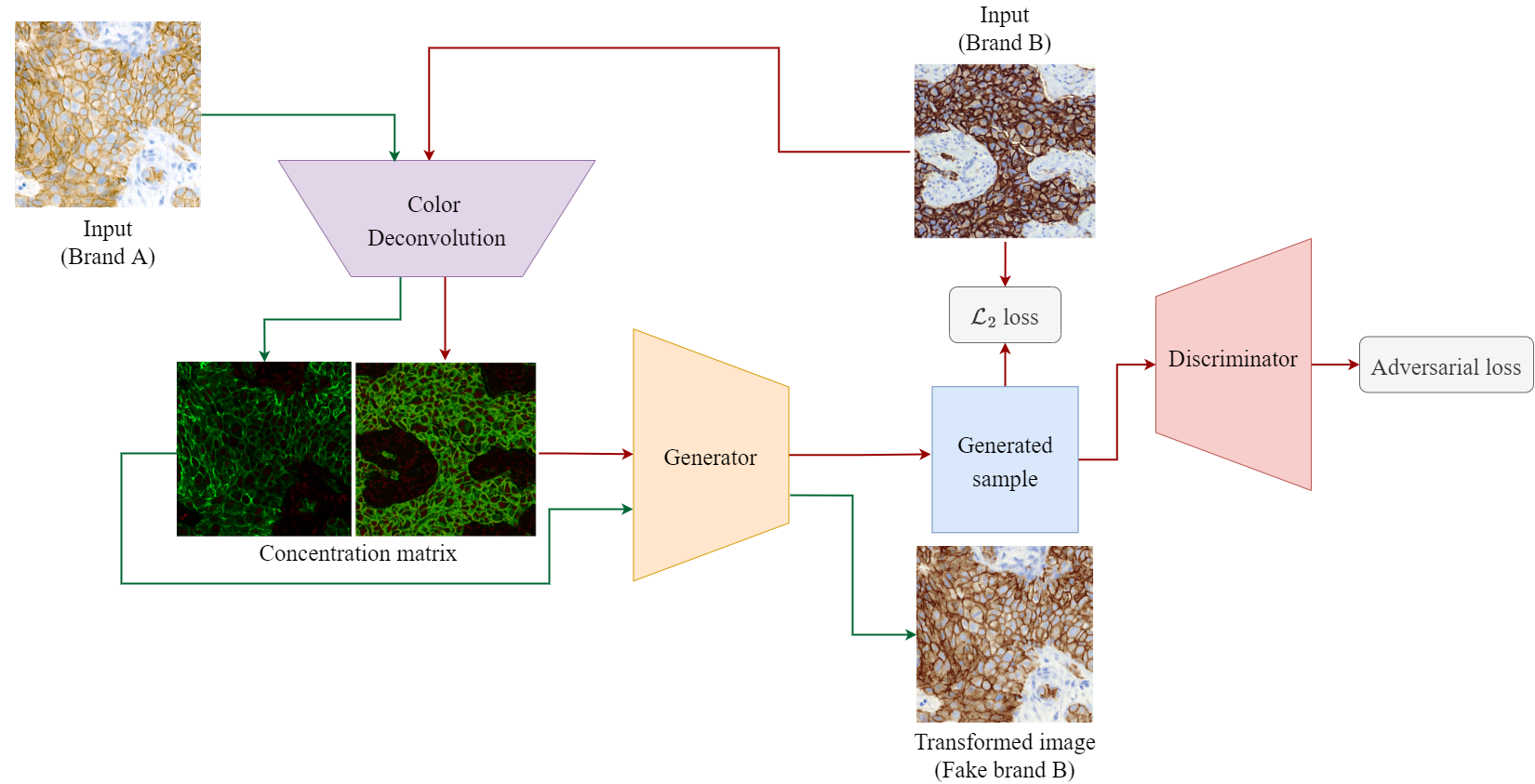}
    \caption{Graphic representation of the training (red arrows) and inference (green arrows) procedure.}
    \label{fig:train-test}
\end{figure*}

As a result of the selection and tiling process, two datasets were generated, one for each domain (stain brand A and stain brand B), each containing 3,000 tiles. These datasets will be used for training and testing of our final model.
\section{Our contribution}\label{sec:our_contr}

Our method combines the Color Deconvolution technique and the Pix2Pix GAN network to perform style transfer in HER2 images. Our main hypothesis is that by computing a domain-independent representation for the stained images, we can recover them in the desired style maintaining the intensity of the membranes to be able to recover the cell classes. 

The proposed architecture is an Encoder-Decoder, where the Encoder receives a HER2 stained tile from a specific hospital (stain brand A) and computes the domain-independent representation of the image, whereas the Decoder recovers the original image from this representation, but in the stain brand B style. Thus, a key challenge is to find an effective method for computing the domain-independent representation. 

For the Encoder, we use the color deconvolution technique to decompose  the images into the color vector matrix $M$ and the concentrations matrix $C$ described in Section \ref{sec:color-dec}.



In the case of HER2 stain, we have 2 different stain channels, one for the brown color (DAB) and one for the blue color (Hematoxylin), and a third channel for residual noise. As domain-independent representation we use the concentrations matrix $C$. To compute this concentrations matrix, a reference color matrix $M_{ref}$ is needed. Using the QuPath application, we extracted the color vectors for each one of our domains. Then, for each image, we compute the concentrations matrix for each one of the stains and the noise channel.

In P\'erez-Bueno F. \textit{et al.} \cite{color_dec_use}, they present an approach that optimizes the reference color vector matrix over the given image, in order to improve the results of the color deconvolution algorithm. We have applied this optimization process over the reference color vector matrices $M_B$ and $M_A$.

For the Decoder, we use the Pix2Pix GAN network, presented by Isola P. \textit{et al.} \cite{pix2pix}. Using this conditional GAN architecture, the Decoder is responsible for recovering the images in the stain brand B style from the representation, which is going to be the concentration matrices of the image.

To perform the translation, the Pix2Pix GAN network is trained only using stain brand B images and their concentration matrices. We feed the Encoder with stain brand B images and, with the computed domain-independent representations, we train the whole conditional GAN architecture. During this procedure, the Decoder learns how to recover images in the stain brand B style just using the information stored in the concentration matrices, which is nearly composed only by domain-independent features. As mentioned in \cite{pix2pix}, this conditional GAN is trained with two different loss functions. First, we train it using the classical adversarial loss of the GANs architectures but, in this case, it also includes a $\mathcal{L}_2$ loss between the original and the generated images (Figure \ref{fig:train-test}).

Once the whole Decoder is trained, to actually perform the style transfer, we just freeze the Generator of the Pix2Pix GAN architecture, and feed the whole model (Encoder and Decoder) using stain brand A images (Figure \ref{fig:train-test}). Hence, using the Color Deconvolution technique, we compute the domain-independent concentration matrices and feed them to the Generator, which recovers the image in the stain brand B style.

We developed two models, combining the optimization process in the computation of the reference color vector matrices in the Color Deconvolution technique \cite{color_dec_use} (suffix WO / O for no optimization / optimization in the model name).

\begin{figure*}[htb]
    \centering
    \includegraphics[scale=0.4]{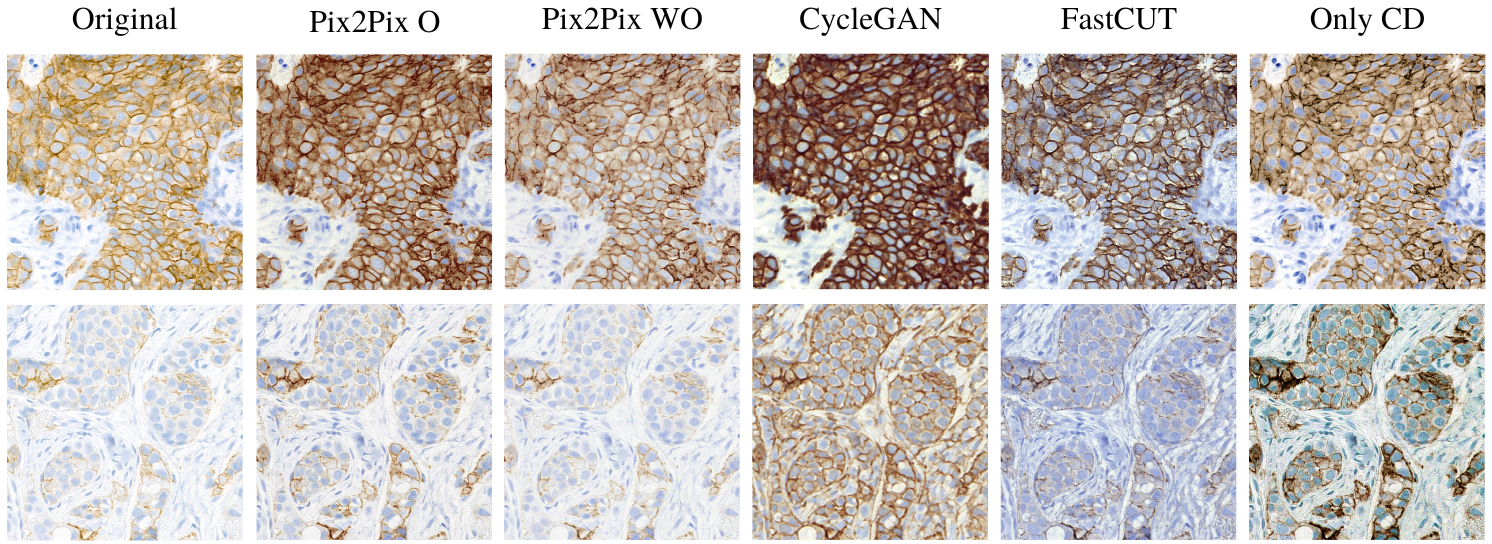}
    \caption{Visualization of original tiles (left) and the obtained translated images for the different models.}
    \label{fig:results}
\end{figure*}

\section{Results}\label{sec:results}

We present the results of our approach to perform the style transfer task between stain brands A and B in HER2 images. As mentioned before, this approach aims to generate realistic images while preserving the HER2 score of the input image. For that reason, we evaluate our results using the Fr\'echet Inception Distance (FID) metric \cite{fid-paper}, to measure the realism of the generated images, and the Weighted F1-Score as a task metric to ensure that the translation maintains the HER2 classes of the cells. We assess our model with these metrics and compare its performance against three baseline models: CycleGAN, FastCUT, and Color Deconvolution (see Table \ref{tab:results}). 

\textbf{FID}. We have used the implementation presented by M. Seitzer \cite{fid}. In our experiments, we found that our proposed combination of Color Deconvolution and Pix2Pix GAN network model had a FID score of 166, which is a good indication of the high quality of the generated images.

\textbf{Weighted F1-Score}. To evaluate the maintenance of cell classes on the generated images, we have computed the class of each cell with a U-Net classification model \cite{her2_mireia} trained with images of the brand B style. We have compared the transformed images classification results with the ground truth classification of the images from stain brand A style.

In terms of the FID metric, our approach outperforms the Color Deconvolution and achieves nearly as good scores as the state-of-the-art style transfer method FastCUT and CycleGAN. Regarding the Weighted F1-Score, our both models outperform the Color Deconvolution baseline as well as the FastCUT and CycleGAN based techniques.

In summary, the experimental results demonstrate that our proposed GAN-based model achieves similar performance than the state-of-the-art models for style transfer while better preserving the structure of cell classes.

\begin{table}[htb]
    \centering
    \begin{tabular}{|r|r|c|c|}
        \hline
        & & \textbf{ $\downarrow$ FID} & \textbf{$\uparrow$ W - F1}\\
        \hline
         \textbf{Pix2Pix + CD} & \textbf{Pix2Pix WO} & \textbf{165.08} & 0.6395 \\
         & \textbf{Pix2Pix O} & 166.61 & \textbf{0.6641}
         \\\hline
         \textbf{CD} & \textbf{Only CD} & 195.79 & 0.3986\\
         \hline
         \hline
         & \textbf{CycleGAN} & 160.52 & 0.5616\\
         & \textbf{FastCUT} & 159.83 & 0.4514 \\
         \hline
    \end{tabular}
    \caption{Results table: All metrics have been computed over a test set of 144 images of stain brand A style}
    \label{tab:results}
\end{table}





\section{Conclusions}\label{sec:discussion}

In this study, we have proposed a novel approach to address the challenge of color variations between HER2-stained histopathological images from different domains. Our method combines the Color Deconvolution technique and the Pix2Pix GAN network to perform a style transfer. 

We have shown that this approach is able to generate images as realistic as those generated by state-of-the-art style transfer models. Moreover, it outperforms the linear approach of using only Color Deconvolution. This indicates that the non-linearity introduced by the Pix2Pix network is effective in solving the style transfer problem in histopathological images. We have used the Weighted F1-score as a task metric to evaluate the translation of cell classes during the style transfer application. Results indicate that our approach is better at maintaining the cell classes than the state-of-the-art models.

Using the proposed translation between stain brands, multicenter projects can perform all the inference with a single, common model for hospitals that may use different HER2 stain brands. Thus, less effort is required in acquiring the data, making our approach a practical and effective solution for improving cancer diagnosis and treatment outcomes. 

To the best of our knowledge, this is the first approach that addresses the color normalization problem in HER2-stained histopathological images, where maintaining the intensity of the membranes and the structure of the image is key to maintain the HER2 score of the image. Overall, this work contributes to the development of more accurate and efficient pathological image analysis tools, which could ultimately improve clinical decision-making and patient outcomes.
\section{Acknowledgments}

This reseach has been partially funded by the project DigiPatICS, from ICS, and the project PID2020-116907RB-100 AIMING, funded by MCIN/ AEI /10.13039/ 501100011033.

\bibliographystyle{IEEEbib}
\bibliography{references}

\end{document}